\documentclass[a4paper,12pt]{article}
\usepackage{graphicx}
\usepackage{amsmath}
\usepackage{amssymb}
\usepackage{epsfig}
\usepackage{xcolor}
\usepackage{hyperref}
\usepackage{cite}
\usepackage{graphicx}
\usepackage{subfigure}
\usepackage{nicefrac, xfrac}
\usepackage[normalem]{ulem}
\newcommand{\nc}{\newcommand}
\nc{\beq}{\begin{equation}} \nc{\eeq}{\end{equation}}
\nc{\beqa}{\begin{eqnarray}} \nc{\eeqa}{\end{eqnarray}}
\nc{\ba}{\begin{array}} \nc{\ea}{\end{array}}

\def\al{\alpha}
\def\be{\beta}

\def\la{\lambda}

\def\pa{\partial}

\def\Ga{\Gamma}

\begin{document}
\begin{center}
{\bf \Large Effective potential in $SO(N)$ symmetric scalar field theories in curved spacetime}
\vspace{1.0cm}

{\bf \large  V.A. Filippov$^{1,a}$, R.M. Iakhibbaev$^{1,b}$  \\[0.3cm] and D. M. Tolkachev$^{1,2,c}$  }

\vspace{0.5cm}
{\it $^1$Bogoliubov Laboratory of Theoretical Physics, Joint Institute for Nuclear Research, 6, Joliot Curie, 141980 Dubna, Russia\\ and \\
$^2$ Stepanov Institute of Physics,
68, Nezavisimosti Ave., 220072, Minsk, Belarus\\}
\vspace{0.5cm}

\abstract{We derive recurrence relations for leading logarithmic all-loop quantum corrections in the case of $SO(N)$ symmetric scalar theory with an arbitrary potential in curved spacetime. On this basis, a system of renormalisation group (RG) equations in the general is obtained approach for the effective potential in the large $N$ limit. As a simple illustration, we analyse the case of power-like potentials in the Jordan frame and discuss their application to inflationary cosmology.}
\\
\textit{E-mails}: {$^a$vafilippov@theor.jinr.ru, $^b$yaxibbaev@jinr.ru, $^c$den3.1415@gmail.com }
\end{center}

\section{Introduction}

\quad The effective potential plays a significant role in studies of vacuum stability and dynamical symmetry breaking \cite{Coleman:1973jx,Jackiw:1974cv, Kastening:1991gv}. The method proposed in \cite{Coleman:1973jx,Jackiw:1974cv} allows one to systematically compute quantum corrections for renormalisable models (such as the $\phi^4$ model and electrodynamics) in an arbitrary loop approximation. The effective potential for the scalar theory $\phi^4$ was further obtained up to two and three loops in flat spacetime \cite{Ford:1991hw, ChungThreeLoop:1998} and one loop in curved spacetime \cite{Buchbinder:1987jf, Buchbinder:2017lnd, Filippov:2025}. An approach developing the study of non-renormalisable interactions was proposed in \cite{Kazakov:2022pkc, Kazakov:2023tii}, where a method for calculating all-loop quantum corrections to the effective potential in the leading logarithmic approximation for an arbitrary potential was presented. This approach was successfully generalised to the case of ordinary scalar models in curved spacetime, as well as to $SO(N)$ symmetric theories in flat space. The aim of this work is to extend previous studies and obtain recurrence relations and the corresponding RG equations in the general approach for the effective potential of $SO(N)$ symmetric scalar field theories with the non-minimal coupling to gravity.
For these purposes, we consider the Lagrangian density of the $SO(N)$ scalar model in curved spacetime:
\beq
\mathcal{L} = -\dfrac{1}{2} R - \dfrac{1}{2} \xi R \phi^2 + \dfrac{1}{2}g_{\mu\nu}\partial^\mu\phi_a\partial^\nu\phi^a - \la V_0(\phi),
\label{lagrangian}
\eeq
where $g_{\mu\nu}$ is the metric tensor, $R$ is the Ricci scalar,  $\la$ is the coupling constant, $ V_0(\phi)$ is the classical potential, $\xi R \phi^2$ is the term of non-minimal interaction of the scalar field with curvature, and $\phi^2 = \phi_a \phi^a$ ($a=1,2,...,N)$, also hereafter we assume $M_{\mathrm{Pl}}=1$.

The paper is organised as follows. In Section 2 we present the Feynman rules, demonstrating the approach to expanding the effective potential into parts dependent on flat and curved backgrounds. We also calculate one- and two-loop corrections up to first order in the curvature $R$. In Section 3, using the Bogoliubov–Parasiuk theorem on the locality of counter terms \cite{BP,Hepp,Zimmermann}, we obtain recurrence relations for the leading divergences and find the corresponding RG equations in the general approach for the effective potential in the large $N$ limit. In Section 4 we consider the power-like potential in the Jordan frame, namely, $\phi^4$ and $\phi^6$. The resulting all-loop effective potentials are studied within the framework of cosmological inflation theory in Section 5. The cosmological parameters for them are computed and compared with the observed Planck 2018 data \cite{Planck:2018jri} and the combined data of Planck-2018, ACT-2025 \cite{ACT:2025blo} and BICEP/Keck 2021 \cite{BICEP:2021xfz} (P-ACT-LB-BK).

\section{Effective potential for $SO(N)$ symmetric interaction in curved spacetime}

\quad The effective potential $V_{eff}(\phi)$ is the part of the effective action $\Gamma[\phi, g_{\mu\nu}]$ that does not contain derivatives of the field $\phi$. So the effective action can be written in the form \cite{Buchbinder:2017lnd}:
\beq
\Ga[\phi,g_{\mu\nu}]=
\int d^4x\sqrt{-g}\left[-V_{eff}(\phi) + \frac{1}{2} Z(\phi) g_{\mu\nu}\pa^\mu\phi_a\pa^\nu\phi^a + ... \right].
\eeq
where the first term corresponds to the effective potential and $Z(\phi)$ is a correction to the wavefunction. The effective potential can be conveniently expanded into the coupling constant $\la$ and split into two parts within the framework of the local-momentum representation \cite{Ishikawa:1983kz, Filippov:2025}:
\beq
V_{eff} (\phi) = \sum_{k=0}^\infty \left(- \dfrac{\lambda}{16\pi^2}\right)^k \left( \lambda \textbf{V}_k(\phi) + \textbf{W}_k(\phi) \right) = \textbf{V}(\phi) + \textbf{W}(\phi).
\label{series}
\eeq
where $\textbf{V}_k$ and $\textbf{W}_k$ are $k$-loop quantum corrections to the effective potential on a flat and curved (in an external gravitational field) spacetime backgrounds, respectively.

\begin{figure}[ht]
 \begin{center}
  \epsfxsize=12cm
 \epsffile{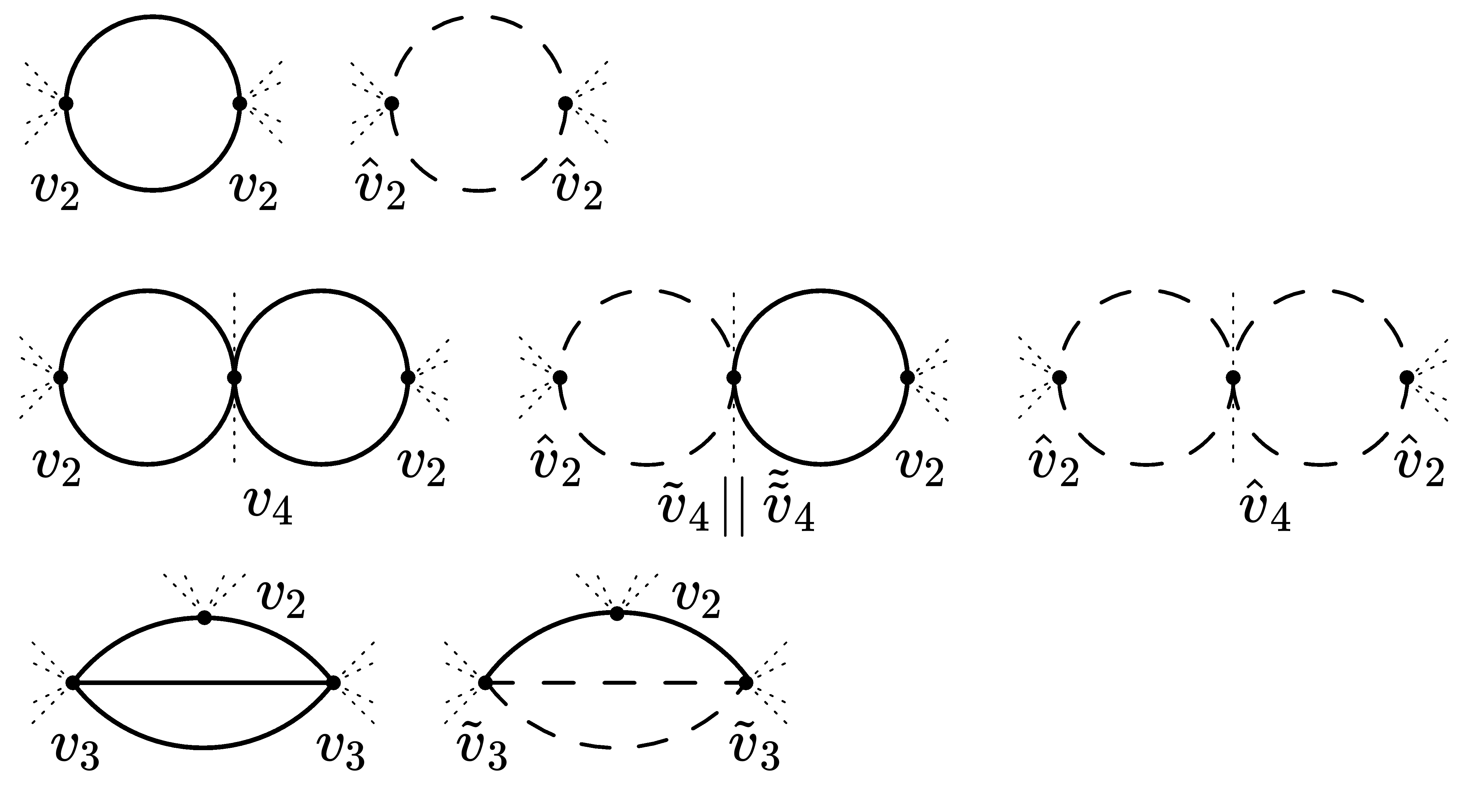}
 \end{center}
 \vspace{-0.2cm}
 \caption{One- and two-loop diagrams giving contributions to $\textbf{V}_k$} 
\label{fig::flat_graphs}
 \end{figure}

We begin by considering the case in flat spacetime. Following the background field splitting \cite{Jackiw:1974cv}, one can obtain the two-point Green's function $G$:
\beq
G_{ab}^{-1}(x,y) = - \left[ \Box \delta_{ab} + \la v_2 \dfrac{\phi_a \phi_b}{\phi^2} + \la \hat{v_2} \left( \delta_{ab} - \dfrac{\phi_a \phi_b}{\phi^2} \right) \right] \delta^4(x-y),
\label{D_inv_ab}
\eeq
where the quantities related to the derivatives of the classical potential $V_0$ are introduced:
\beq
v_2 = \dfrac{\partial^2 V_0(\phi)}{\partial \phi \partial \phi}, \quad \hat{v}_2 = 2 \dfrac{\partial V_0(\phi)}{\partial (\phi^2)}.
\eeq
Further, it is convenient to switch to the momentum representation and take the inverse expression:
\beq
G_{ab}(p) = \dfrac{1}{p^2 - m_1^2} \dfrac{\phi_a \phi_b}{\phi^2} +  \dfrac{1}{p^2 - m_2^2} \left( \delta_{ab} - \dfrac{\phi_a \phi_b}{\phi^2} \right).
\eeq
It can be seen how the propagator factors into a diagonal part with mass $m_1^2 = \la v_2$ and a non-diagonal part contributing with $\hat{N}=N-1$ and mass $m_2^2= \la \hat{v}_2$. The one- and two-loop one-particle irreducible diagrams contributing to $\textbf{V}_k$ are shown in Fig. \ref{fig::flat_graphs}. 

In the formalism under consideration, each vertex is assigned certain factors $v_n, \hat{v}_n$ or $\tilde{v}_n$, where the index $n$ determines the number of quantum field lines entering into the vertex. Thus, for the diagonal case:
\beq
v_n = \dfrac{\partial^n V_0(\phi)}{\partial \phi^n},
\eeq
for non-diagonal terms:
\beq
\hat{v}_2 = 2 \dfrac{\partial V_0(\phi)}{\partial (\phi^2)}, \quad \hat{v}_4 = 2 \dfrac{\partial }{\partial (\phi^2)} \cdot 2 \dfrac{\partial V_0(\phi)}{\partial (\phi^2)},  \quad \text{...}
\eeq
For the vertices with mixed quantum lines, the following factors are introduced:
 \beq
 \tilde{v}_3 = 2 \dfrac{\partial }{\partial (\phi^2)} \cdot\dfrac{\partial V_0(\phi)}{\partial \phi}, \quad \tilde{v}_4 = 2 \dfrac{\partial }{\partial (\phi^2)} \cdot\dfrac{\partial^2 V_0(\phi)}{\partial \phi^2}, \quad 
  \tilde{\tilde{v}}_4 = \dfrac{\partial^2 }{\partial \phi^2} \cdot 2 \dfrac{\partial V_0(\phi)}{\partial (\phi^2)},
 \quad \text{...}
 \eeq
Note that the operators at the vertices with four quantum lines are non-commutative, which is clear from the form of the corresponding derivatives.
 
Taking into account the Feynman rules, analytical expressions can be presented for the diagrams shown in Fig. \ref{fig::flat_graphs}. The one-loop quantum corrections $\textbf{V}_1$ in dimensional regularisation with $d=4-2\varepsilon$ take the following form:
 \beq
 \textbf{V}_1=\frac{1}{4} v_2^2 \left( \frac{1}{\varepsilon} + \log\left( \frac{\mu^2}{m^2_1}\right)\right) + \frac{\hat{N}}{4}\hat{v}_2^2\left( \frac{1}{\varepsilon} + \log\left( \frac{\mu^2}{m^2_2} \right) \right).
 \label{VV_1}
 \eeq
Expression \eqref{VV_1} contains both leading divergences ${1}/{\varepsilon}$ and leading logarithms with masses corresponding to various propagators. It is important to note that the coincidence of the coefficients of the leading logarithms and leading divergences is a general property for any UV-divergent Feynman loop graph. In view of this observation, in the present work we restrict ourselves to the leading logarithmic approximation of the effective potential. Therefore, to study the corrections after renormalisation, we use the following substitution:
\beq
\frac{1}{\varepsilon} \rightarrow \log\left( \frac{\mu^2}{m^2_i}\right)
\label{epsLog}
\eeq
We further introduce the notation $\Delta \textbf{V}_k$ for the singular parts of the diagrams, while by $\textbf{V}_k$ we denote the leading logarithmic $k$-loop corrections to the effective potential.
The two-loop diagrams in the flat case take the following compact form:
\beq
\Delta \textbf{V}_2 = \dfrac{1}{8 \varepsilon^2} \left( v^2_2 v_4 +  v_2 v_3^2 \right) + \dfrac{\hat{N}}{8 \varepsilon^2} \left( v_2 \hat{v}_2 \tilde{v}_4 + v_2 \hat{v}_2 \tilde{\tilde{v}}_4 + v_2 \tilde{v}_3^2 \right) + \dfrac{\hat{N}^2}{8 \varepsilon^2} \hat{v}^2_2 \hat{v}_4.
\label{VV_2}
\eeq
From these, by performing the substitution \eqref{epsLog} we can obtain an expression for $\textbf{V}_k$.
The above expressions can be easily written down for the special case of the theory with a quartic interaction potential $(\phi_a \phi_a)^2/4!$:
\beq
 \textbf{V}_1=\frac{1}{16} \phi^4 \log\left( \frac{\mu^2}{m^2_1}\right) + \frac{\hat{N}}{144}\phi^4 \log\left( \frac{\mu^2}{m^2_2} \right),
\eeq
\beq
\textbf{V}_2 = \dfrac{3}{32} \phi^4 \log^2\left( \frac{\mu^2}{m^2_1}\right) + \dfrac{\hat{N}}{48} \phi^4 \log\left( \frac{\mu^2}{m^2_1}\right) \log\left( \frac{\mu^2}{m^2_2}\right) + \dfrac{\hat{N}^2}{864} \phi^4 \log^2\left( \frac{\mu^2}{m^2_2}\right).
\eeq
These results are known (see, e.g., Ref \cite{ChungThreeLoop:1998}) and are usually subject to renormalisation conditions of the form $$ \frac{\partial^4}{\partial \phi^4}\left(\lambda \textbf{V}(\phi)\right)\bigg|_{\phi_c=M_*}=\lambda,$$
where $M_*$ is some arbitrary scale on which the observables do not depend \cite{Coleman:1973jx}.  It is worth reminding here that the coupling constant $\la$ was taken out of the initial potential and loop corrections for convenience, see expression \eqref{series}.

To extend our analysis to curved spacetime, we employ the local-momentum representation\cite{Bunch:1979uk, Birrell:1982ix, Petrov:1969}, which is based on the introduction of Riemann normal coordinates $y^\mu=x^\mu-x'^\mu$:
\beq
g_{\al\be}(x) = \eta_{\al\be}
-\frac13\,R_{\al\mu\be\nu}(x^\prime)\,y^\mu \,y^\nu\ + \ldots,
\eeq
where the expansion is performed around the point $x^\prime$ at which $g_{\al\be}(x^{\prime})=\eta_{\al\be}$.
Here the dots denote higher-order terms in the scalar curvature $R$, which is beyond the scope of the present work. This approach allows one to use the usual Feynman rules described earlier with the addition of a classical gravitational field background, while preserving the overall covariance of the results. 

Thus, to first order in the curvature $R$, we can obtain modified propagators in momentum space \cite{Bunch:1979uk,Sobreira:2011ep,Iakhibbaev:2024fjf}:
\beq
G_{ab}^{(1)}(p) = \left( \dfrac{1}{p^2 - m_1^2} + \dfrac{\hat{\xi} R}{(p^2 - m_1^2)^2} \right) \dfrac{\phi_a \phi_b}{\phi^2},
\label{G_1}
\eeq

\beq
G_{ab}^{(2)}(p) = \left( \dfrac{1}{p^2 - m_2^2} + \dfrac{\hat{\xi}R}{(p^2 - m_2^2)^2} \right) \left( 
\delta_{ab} - \dfrac{\phi_a \phi_b}{\phi^2} \right),
\label{G_2}
\eeq
where $\hat{\xi} = \xi - \dfrac{1}{6}$. Denote the quantum corrections to the effective potential in the external gravitational field as $\textbf{W}_k$.

\begin{figure}[ht]
 \begin{center}
  \epsfxsize=9cm
 \epsffile{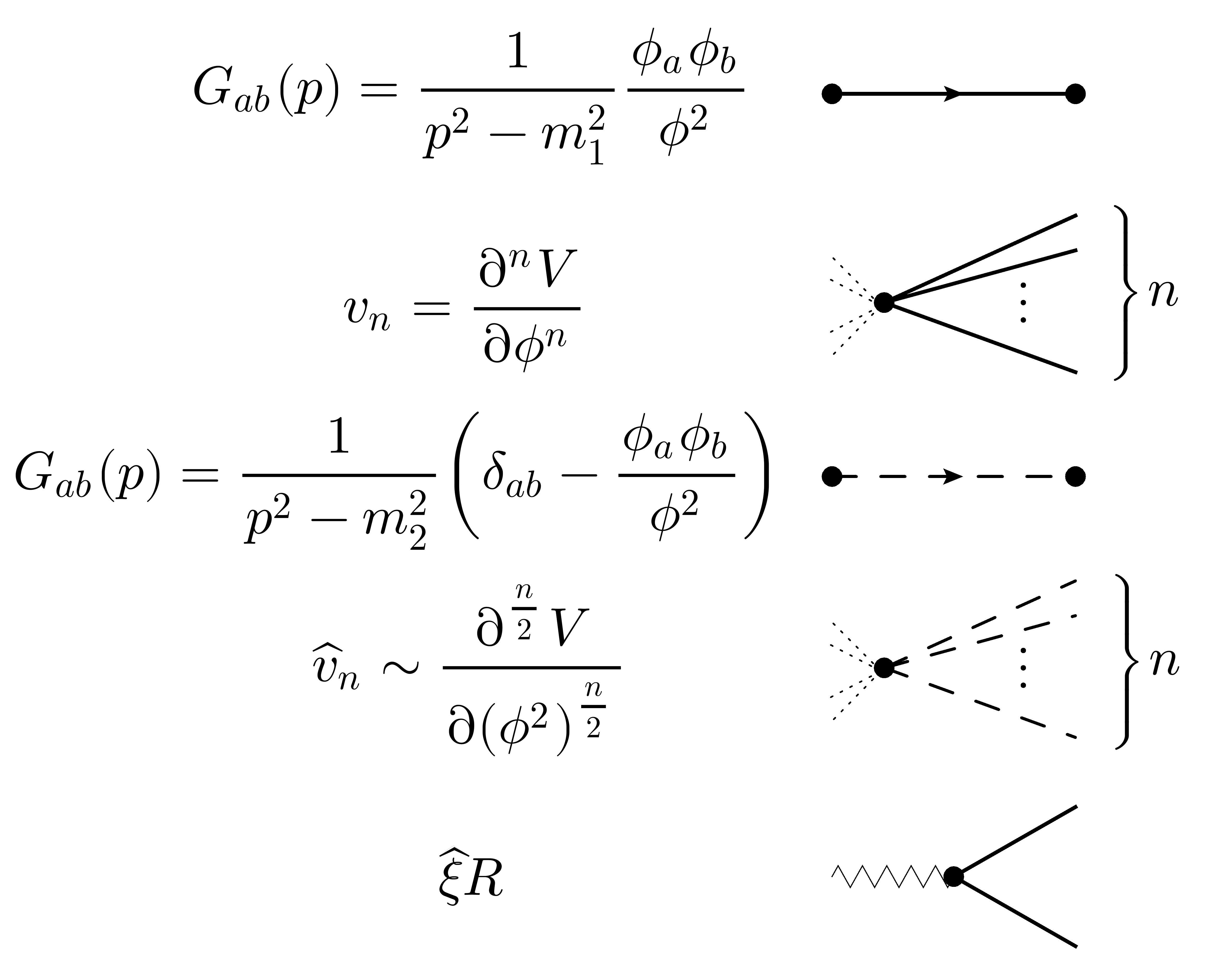}
 \end{center}
 \vspace{-0.6cm}
 \caption{Modified Feynman rules
 } 
\label{fig::F_rules}
 \end{figure}

\begin{figure}[ht]
 \begin{center}
  \epsfxsize=12cm
 \epsffile{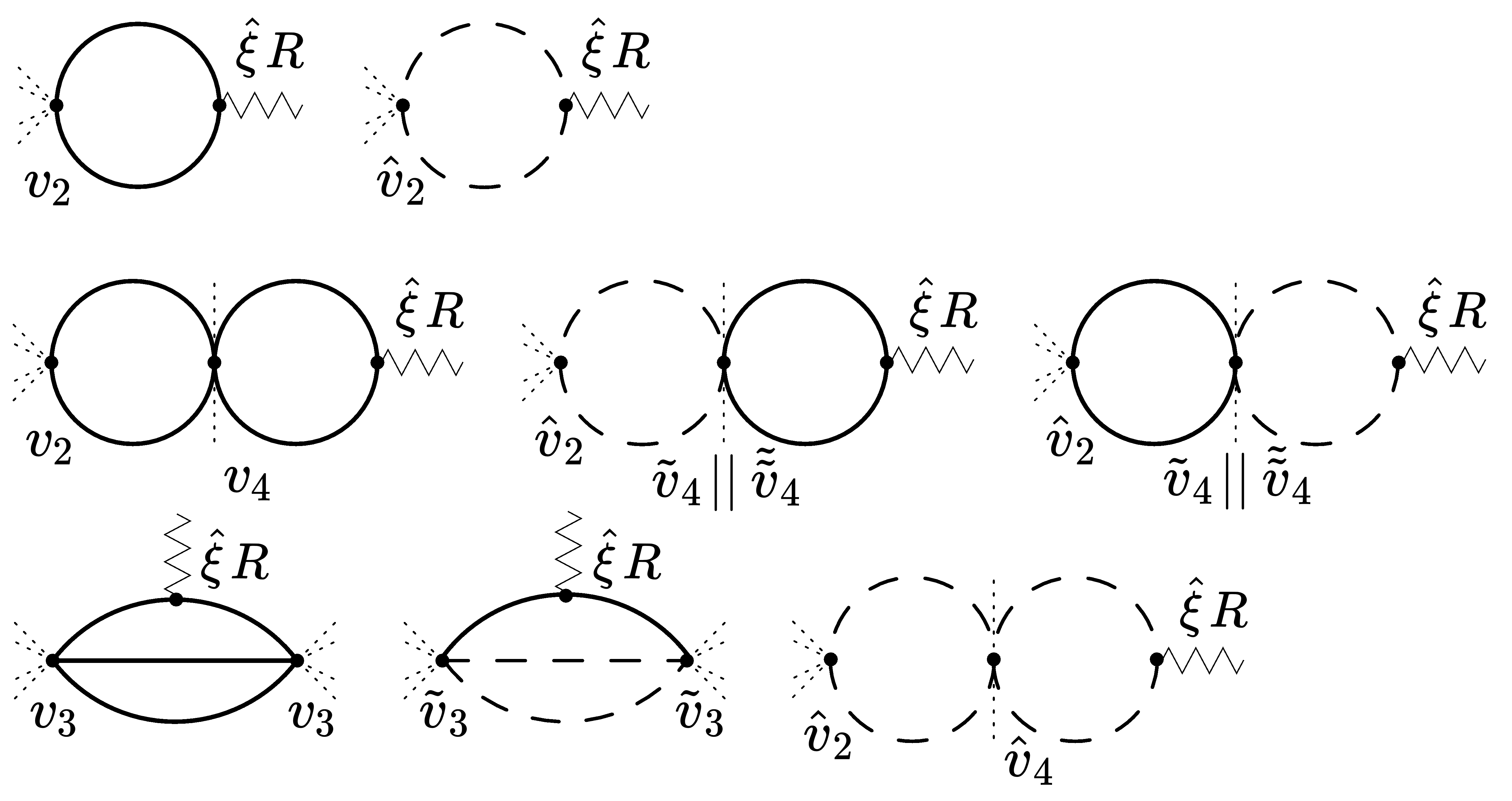}
 \end{center}
 \vspace{-0.6cm}
 \caption{One- and two-loop diagrams giving contributions to $\textbf{W}_k$} 
\label{fig::curved_graphs}
 \end{figure}

The one-loop corrections calculated in the standard way as $\text{Tr}\log G(x,x^\prime)$ can be written separately for the contributions from \eqref{G_1} and from \eqref{G_2}:
\beq
V_1=\textbf{V}_1+\textbf{W}_1,
\eeq
where $\textbf{V}_1$ is defined by expression \eqref{VV_1}, and the contribution to the one-loop correction on a curved background is given by $\textbf{W}_1$:
 \beq
 \textbf{W}_1=\frac{1}{2} v_2 \hat{\xi} R \left( \frac{1}{\varepsilon} + \log\left( \frac{\mu^2}{m^2_1}\right)\right) + \frac{\hat{N}}{2}\hat{v}_2 \hat{\xi} R\left( \frac{1}{\varepsilon} + \log\left( \frac{\mu^2}{m^2_2} \right) \right).
 \label{WW_1}
 \eeq
Using expressions \eqref{VV_1}, \eqref{VV_2} and \eqref{WW_1}, we then derive the modified Feynman rules, which are shown in Fig. \ref{fig::F_rules}.
The one- and two-loop corrections contributing to the effective potential on curved spacetime, denoted as $\textbf{W}$, are shown in Fig. \ref{fig::curved_graphs}. A comparison of expressions \eqref{WW_1} and \eqref{VV_1} demonstrates a difference in symmetry factors arising from the loss of symmetry due to the inclusion of the additional vertex associated with the curvature $R$. We also provide an expression for the two-loop contribution $\mathbf{W}_2$ to the effective potential:

\beq
\begin{split}
\Delta \textbf{W}_2 &= \dfrac{1}{8 \varepsilon^2} \left( 2 v_2 v_4 \hat{\xi} R + v_3^2 \hat{\xi} R \right)+ \dfrac{\hat{N}^2 }{4 \varepsilon^2} \hat{v}_2 \hat{v}_4 \hat{\xi} R + \\
&+ \dfrac{\hat{N}}{8 \varepsilon^2} \left( v_2 \tilde{v}_4 \hat{\xi} R + v_2 \tilde{\tilde{v}}_4 \hat{\xi} R + \hat{v}_2 \tilde{v}_4 \hat{\xi} R + \hat{v}_2 \tilde{\tilde{v}}_4 \hat{\xi} R + \tilde{v}_3^2 \hat{\xi} R \right).
\label{WW_2}
\end{split}
\eeq

Let us again consider the theory with a quartic potential $(\phi_a \phi_a)^2/4!$. The one-loop correction $\textbf{W}_1$ takes the form:
\beq
 \textbf{W}_1=\frac{1}{4} \phi^2 \hat{\xi} R \log\left( \frac{\mu^2}{m^2_1}\right) + \frac{\hat{N}}{12} \phi^2 \hat{\xi} R \log\left( \frac{\mu^2}{m^2_2} \right),
\eeq
which corresponds to \cite{Buchbinder:2017lnd}. The two-loop contribution $\textbf{W}_2$ is given by:
\beq
\textbf{W}_2 = \dfrac{1}{4} \phi^2 \hat{\xi} R \log^2\left( \frac{\mu^2}{m^2_1}\right) + \dfrac{\hat{N}}{8} \phi^2 \hat{\xi} R \log\left( \frac{\mu^2}{m^2_1}\right) \log\left( \frac{\mu^2}{m^2_2}\right) + \dfrac{\hat{N}^2}{72} \phi^2 \hat{\xi} R \log^2\left( \frac{\mu^2}{m^2_2}\right).
\eeq
To proceed, we should analyse the obtained loop contributions to construct recurrence relations for the leading poles.

\section{Recurrence relations and  RG equations in the general approach}

\begin{figure}[ht]
 \begin{center}
  \epsfxsize=14cm
 \epsffile{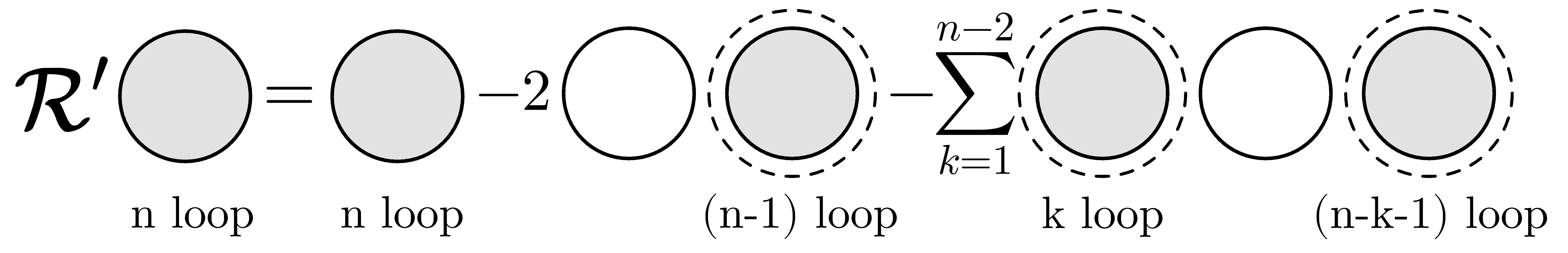}
 \end{center}
 \vspace{-0.4cm}
 \caption{$\mathcal{R}^{\prime}$ operation for leading divergences of an $n$-loop diagram} 
\label{fig::R_oper}
 \end{figure}

\quad In this section, we analyse the coefficients at leading poles in multi-loop diagrams to establish relations between them, and then we perform the substitution \eqref{epsLog} to obtain the desired effective potential.
Recall that the ${\mathcal{R}}$ operation without the last subtraction of the complete graph (i.e., ${\mathcal{R}^{\prime}} $ operation) \cite{BogoliubovBook}, being applied to an $n$-loop diagram, eliminates UV divergences from the first up to $(n-1)$-loop divergent subgraphs. The action of $\mathcal{R}^{\prime}$ on an arbitrary diagram $G$ can be written as follows:
\beq
\mathcal{R}^{\prime} G =\left( 1 - \sum_{\gamma}\left( \mathcal{K}\mathcal{R}^{\prime} \right)_{\gamma} + \sum_{\gamma, \gamma^{\prime}} \left( \mathcal{K}\mathcal{R}^{\prime} \right)_{\gamma} \left( \mathcal{K}\mathcal{R}^{\prime} \right)_{\gamma^{\prime}} - ... \right) G,
\eeq
where each term $\left( \mathcal{K}\mathcal{R}^{\prime} \right)_{\gamma}$ acts on $G$ as:
\beq
\left( \mathcal{K}\mathcal{R}^{\prime} \right)_{\gamma} G = \mathcal{K}\mathcal{R}^{\prime}(\gamma) \circ \nicefrac{G}{\gamma},
\eeq
In the expressions above, $\mathcal{K}_\gamma$ extracts the singular part of the subgraph $\gamma$, $\nicefrac{G}{\gamma}$ represents the diagram $G$ with the subgraph $\gamma$ shrunk to a point, and the $\circ$ operation denotes the insertion of the counter term of the subgraph into the reduced graph $\nicefrac{G}{\gamma}$. The remaining $n$-loop leading divergence $A_n^{(n)}$ is exactly what we need. The leading divergence $A_n^{(n)}$ must be local according to the Bogoliubov-Parasiuk theorem \cite{BP,Hepp,Zimmermann}. This requirement results in a relation between the leading divergence in $n$-loop $A_n^{(n)}$ and the contribution from the one-loop diagram $A_n^{(1)}$ obtained after subtracting the $(n-1)$-loop counter term:
\beq
A_n^{(n)} = \dfrac{(-1)^{n+1}}{n} A_n^{(1)}.
\eeq
The action of the $\mathcal{R}^{\prime}$ operation is schematically illustrated in Fig. \ref{fig::R_oper}.

Following the approach of \cite{Kazakov:2022pkc, Filippov:2025}, recurrence relations can be obtained for the coefficients at the leading poles of an arbitrary diagram:
\beq
n \Delta \textbf{V}_n = \dfrac{1}{4} \sum_{k=0}^{n-1} \left( \dfrac{\partial^2 \Delta \textbf{V}_k}{\partial \phi^2} \cdot \dfrac{\partial^2 \Delta \textbf{V}_{n-k-1}}{\partial \phi^2} + 4 \hat{N} \dfrac{\partial \Delta \textbf{V}_k}{\partial (\phi^2)} \cdot \dfrac{\partial \Delta \textbf{V}_{n-k-1}}{\partial (\phi^2)} \right),
\label{rec_V}
\eeq

\beq
n \Delta \textbf{W}_n = \dfrac{1}{2} \sum_{k=0}^{n-1} \left( \dfrac{\partial^2 \Delta \textbf{W}_k}{\partial \phi^2} \cdot \dfrac{\partial^2 \Delta \textbf{V}_{n-k-1}}{\partial \phi^2} + 4 \hat{N} \dfrac{\partial \Delta \textbf{W}_k}{\partial (\phi^2)} \cdot \dfrac{\partial \Delta \textbf{V}_{n-k-1}}{\partial (\phi^2)} \right),
\label{rec_W}
\eeq
where $n \geq 2$.
It is also convenient to introduce functions summing leading divergences:
\beq
\Sigma_\la (z,\phi) = \sum_{n=0}^\infty (-z)^n \Delta \textbf{V}_n, ~~ \Sigma_\xi (z,\phi) = \sum_{n=0}^\infty (-z)^n \Delta \textbf{W}_n,
\eeq
where $z=\frac{\lambda}{16 \pi^2\varepsilon}$.
Then the recurrence relations \eqref{rec_V} and \eqref{rec_W} in the large $N$ approximation can be converted into a system of equations for $\Sigma_\la$ and $\Sigma_\xi (z,\phi)$:
\begin{equation}
\dfrac{\partial\Sigma_{\la}}{\partial z} = - N \left( \dfrac{\partial \Sigma_{\la}}{\partial (\phi^2)} \right)^2,
\label{VRG}
\end{equation}
\begin{equation}
\dfrac{\partial\Sigma_\xi}{\partial z} = -2 N \dfrac{\partial\Sigma_\xi}{\partial (\phi^2)} \cdot \dfrac{\partial\Sigma_\la}{\partial (\phi^2)},
\label{WRG}
\end{equation}
with the initial conditions $\Sigma_\la(0,\phi) = V_0(\phi)$ and $\Sigma_\xi(0,\phi) = \frac{1}{2}\hat{\xi}R \phi^2$.  Expressions \eqref{VRG} and \eqref{WRG} are the desired RG equations in the general approach, valid for an arbitrary classical potential $V_0(\phi)$. 

Finally, the effective potential can be represented using the substitution \eqref{epsLog} as follows:
\beq
V_{eff}=\lambda \Sigma_\la(z,\phi) + \left(\frac{1}{12} R\phi^2 + ~\Sigma_\xi(z,\phi)\right)\bigg|_{z \rightarrow -\frac{\lambda}{16\pi^2}\log(\lambda \hat{v}_2/\mu^2)}
\label{V_eff_result}
\eeq
We use explicit examples to verify our results and analyse the behaviour of effective potentials in renormalisable and non-renormalisable models in the following section.

\section{Power-like potentials}

\quad Let us consider the general case of a power-like potential of the form:
\beq
V_0(\phi) = \dfrac{(\phi^2)^{\frac{p}{2}}}{p!}.
\label{V_0}
\eeq
In this case, it is convenient to use an ansatz expressed in terms of dimensionless variables:
\beq
\begin{split}
&\Sigma_\la(z,\phi) = \dfrac{\phi^p}{p!}f_1(x),\\
&\Sigma_\xi(z,\phi) = \hat{\xi} R \dfrac{\phi^2}{2!}f_2(x)
\end{split}
\label{Sigma_f}
\eeq
where $x=z (\phi^2)^{\frac{p}{2}-2}$. After substituting this ansatz into RG equations in the general approach, we obtain a system of nonlinear ordinary differential equations for the functions $f_1(x)$ and $f_2(x)$:
\beq
f_1'(x) = -\dfrac{N}{4 p!} \left[(p-4) x f_1'(x)+p f_1(x)\right]^2,
\label{eq_f_1}
\eeq
\beq
f_2'(x) = \frac{N}{2 p!} \left((p-4) x f_1'(x)+p f_1(x)\right) \left((p-4) x f_2'(x)+ 2 f_2(x)\right)
\label{eq_f_2}
\eeq
with the initial conditions:
\beq
f_1(0) = 1,\quad f_2(0) = 1.
\eeq
From the requirement, that the expansion parameter $x$ be small and from the leading logarithmic approximation, we establish a constraint on the coupling constant $\la$ and the transmutation parameter $\mu$:
\beq
\lambda^{2} \mu^{2(p-4)} < \dfrac{ (16 \pi^2)^{p-2} }{ [(p-2)!]^{p-4} }
\eeq
Further, we consider definite values of $p$ for the system (\ref{eq_f_1}-\ref{eq_f_2}).

\subsection{Renormalisable $p=4$ case}

\quad For the case of a renormalised power-like potential with $p=4$ in four-dimensional space, the RG equations in the general approach \eqref{eq_f_1}, \eqref{eq_f_2} are significantly simplified, taking the form:
\beq
f_1'(x) = -\dfrac{N}{6} f_1^2(x),
\eeq
\beq
f_2'(x) = -\dfrac{N}{6} f_1(x) f_2(x).
\eeq
The solution to the equations is a geometric progression of the form:
\beq
f_1(x) \equiv f_2(x) = \dfrac{1}{1 + \frac{N}{6} x}
\eeq
with the Landau pole at $x = - \frac{6}{N}$.

\begin{figure}[ht]
    \begin{center}
    \begin{minipage}{0.45\textwidth}
            \epsfxsize=8.5cm
            \epsffile{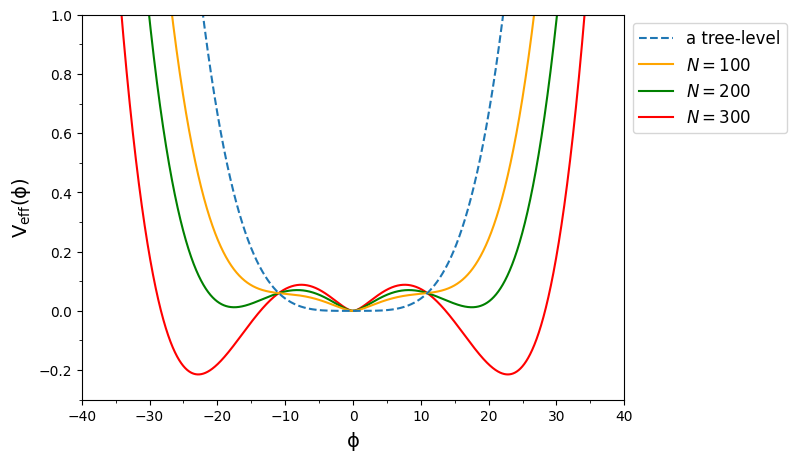}
            \label{fig::Veff_p4_RC1}
        \end{minipage}
        \hfill
        \begin{minipage}{0.45\textwidth}
            \epsfxsize=8.5cm
            \epsffile{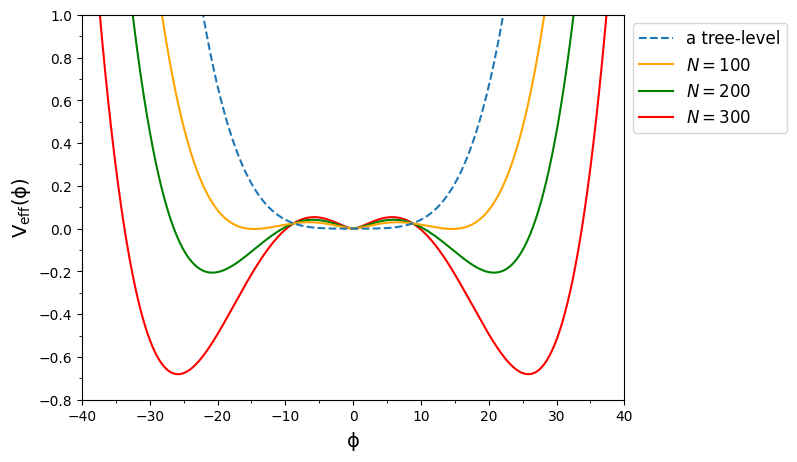}
            \label{fig::Veff_p4_RC2}
        \end{minipage}
    \end{center}
    \vspace{-0.9cm}
    \caption{Effective potential of the model with $p=4$ for $\xi = 0$, various numbers of fields $N$ and curvature values $R = \{R_{C1}, R_{C2}\}$. Critical curvature values $R_{C1}\simeq 50 \cdot \mu^2, R_{C2}\simeq 70 \cdot \mu^2$ are selected for the effective potential with $N=100$.}
    \label{fig::Veff_p4_RC_12}
\end{figure}

The effective potential for various values of $N$ and curvature $R$ is shown in Fig.\ref{fig::Veff_p4_RC_12}. It can be seen that as the number of fields $N$ increases, the additional minima become lower: the value of the effective potential becomes smaller at these minima. Similar to the work \cite{Ishikawa:1983kz}, two positive critical values of the curvature $R_{C1}$ and $R_{C2}$ are considered. Thus, the curvature value $R_{C1}$ is responsible for the appearance of an additional minimum in the effective potential due to quantum corrections, whereas $R_{C2}$ is responsible for the transition of the additional minimum from local to global. For illustrative purposes, the values of $R_{C1}$ and $R_{C2}$ were selected for the effective potential with $N=100$.

\begin{figure}[ht]
 \begin{center}
  \epsfxsize=10cm
 \epsffile{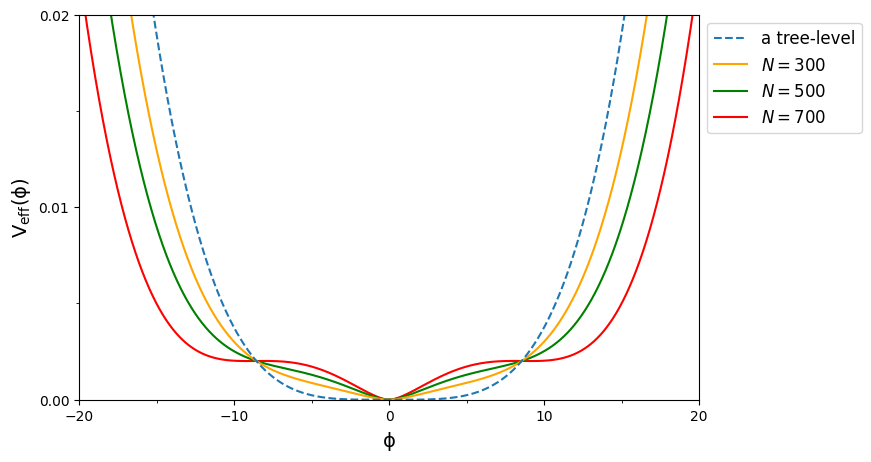}
 \end{center}
 \vspace{-0.4cm}
 \caption{Dependence of the effective potential on various $N$ values for reaching a locally flat plateau at $N=700$} 
\label{fig::Veff_p4_PBHs}
 \end{figure}

Of particular interest is the transition case with the appearance of an additional minimum in the effective potential, in which a flat plateau is formed locally in a natural way. By varying the parameters, in particular the number of fields $N$, one can tune the extension of this flat section. The effective potential for various values of $N$ is shown in Fig.\ref{fig::Veff_p4_PBHs}, where one can see a smooth transition from the classical potential to the effective potential with $N=700$, which has a flat plateau. This regime can be useful in analysis of primordial black holes formation during cosmological inflation \cite{Ivanov:1994, Ezquiaga:2018, Ballesteros:2018, Frolovsky:2023}.  These hypothetical objects are considered as one of the main candidates for the dark matter source in the Universe. In this context, it would be interesting to investigate the behaviour of inflationary potentials within the framework of the problem of primordial black hole formation.

\subsection{$p=6$ case}
\begin{figure}[ht]
    \begin{center}
    \begin{minipage}{0.45\textwidth}
            \epsfxsize=7.0cm
            \epsffile{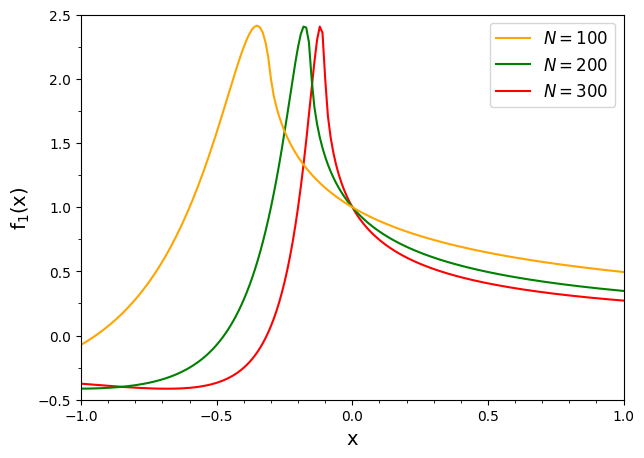}
            \label{fig::f_1}
        \end{minipage}
        \hfill
        \begin{minipage}{0.45\textwidth}
            \epsfxsize=7.0cm
            \epsffile{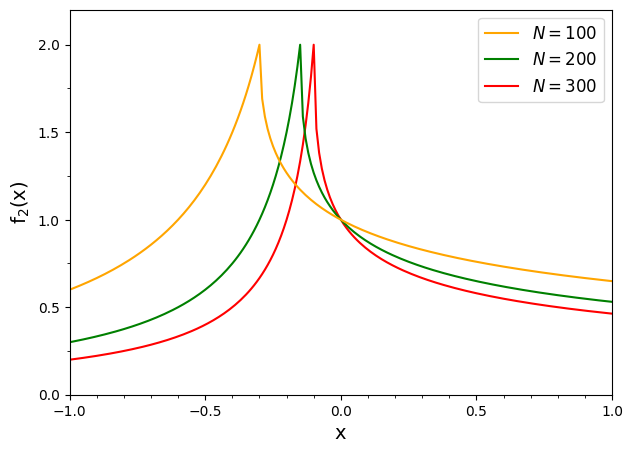}
            \label{fig::f_2}
        \end{minipage}
    \end{center}
    \vspace{-0.9cm}
    \caption{Solutions $f_1(x)$ and $f_2(x)$ for various numbers of fields $N$}
    \label{fig::f_1_2}
\end{figure}

\quad For the power-like potential with $p=6$, analytical solutions can be given for $f_1(x)$ \cite{Iakhibbaev:2024fjf} and $f_2(x)$:
\beq
f_1(x) = 30 \cdot (5!)^2 \frac{1 + \frac{1}{20} N x \left(1+\frac{N x}{120}\right)- \left(1+\frac{N x}{30}\right)^{\frac{3}{2}}}{N^3 x^3}
\eeq
\beq
f_2(x) = 60 \sqrt{2}  \left(\frac{1+\frac{N
   x}{60}-\sqrt{1 + \frac{N x}{30}}}{N^2
   x^2}\right)^{1/2}
\eeq
It follows that the solutions acquire an imaginary part for $x<-\frac{30}{N}$. The solutions $f_1(x)$ and $f_2(x)$ for various $N$ are shown in Fig. \ref{fig::f_1_2}. As $x \rightarrow \infty$, both functions tend to zero, which implies that they remain finite once the regularisation is removed.

\begin{figure}[ht]
    \begin{center}
    \begin{minipage}{0.45\textwidth}
            \epsfxsize=8.5cm
            \epsffile{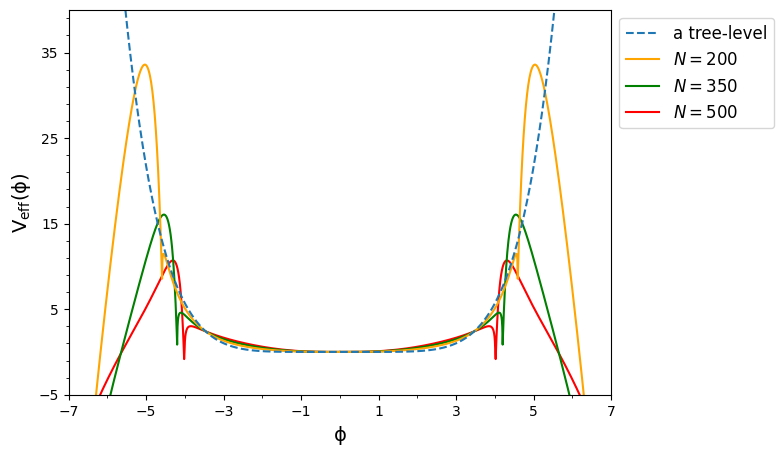}
            \label{fig::Veff_p6_R_n}
        \end{minipage}
        \hfill
        \begin{minipage}{0.45\textwidth}
            \epsfxsize=8.5cm
            \epsffile{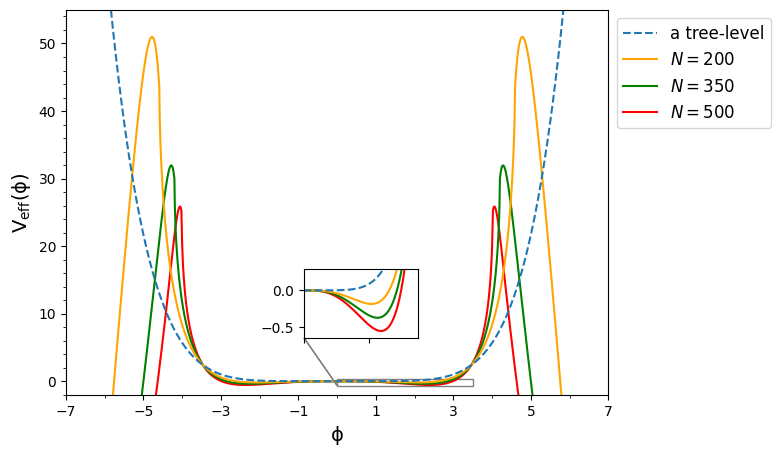}
            \label{fig::Veff_p6_R_p}
        \end{minipage}
    \end{center}
    \vspace{-0.9cm}
    \caption{Effective potential of the model with $p=6$ for $\xi = 0$, various numbers of fields $N$ and curvature values $R$ equal to $R \sim -10 \mu^2$ and $R \sim 10 \mu^2$, respectively. Parameters for illustrative purposes are chosen as $\la \sim 1$, $\mu \sim 1$.}
    \label{fig::Veff_p6_R_n_p}
\end{figure}

It is also possible to obtain the effective potential using expressions \eqref{V_eff_result} and \eqref{Sigma_f}. The result is shown in Fig. \ref{fig::Veff_p6_R_n_p}. The figure shows that as the number of fields $N$ increases, additional minima of the effective potential become more significant.

\begin{figure}[ht]
    \begin{center}
    \begin{minipage}{0.45\textwidth}
            \epsfxsize=8.5cm
            \epsffile{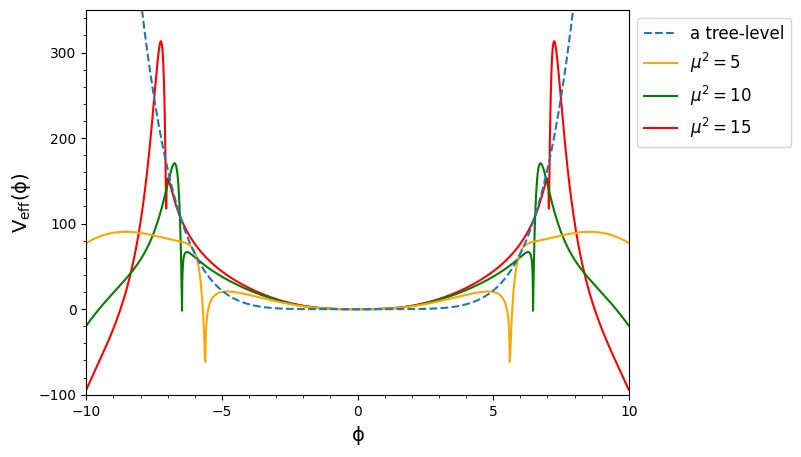}
            \label{fig::Veff_p6_p_Mu}
        \end{minipage}
        \hfill
        \begin{minipage}{0.45\textwidth}
            \epsfxsize=8.5cm
            \epsffile{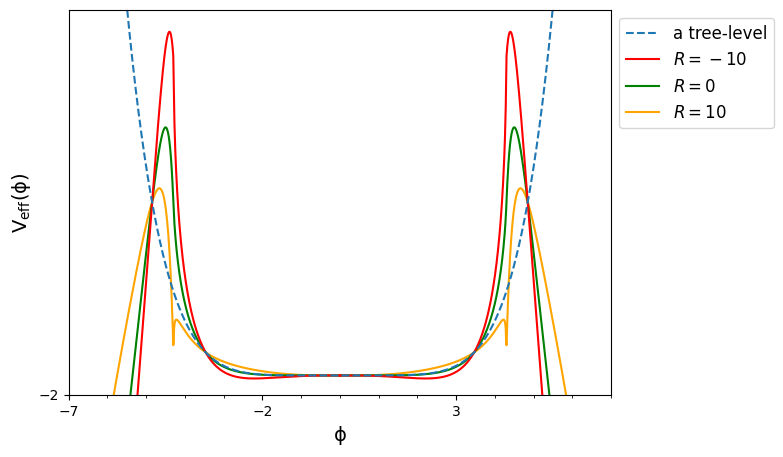}
            \label{fig::Veff_p6_p_R}
        \end{minipage}
    \end{center}
    \vspace{-0.9cm}
    \caption{Effective potential of the model with $p=6$ for $\xi =0$, $N=200$,  $\la \sim 1$ with various values of $\mu$ and fixed $R \sim 60$ (left), and also various values of curvature $R$ and fixed $\mu \sim 1$ (right).}
    \label{fig::Veff_p6_p_mu_R}
\end{figure}

Figure \ref{fig::Veff_p6_p_mu_R} depicts the effective potentials for fixed $N=200$, various values of transmutation parameter $\mu$, and curvature $R$. It can be observed that as $\mu$ increases, additional minima of the potential are lifted. In the right figure, one can follow the qualitative transition in the behaviour of the potential from positive to negative values curvature $R$.  
One can notice that in the case of negative values of $R$ near zero, typical additional minima are formed, and in the case of positive values of curvature, they take on a different form due to the shape of the function $f_2(x)$.

\section{Applications in inflationary cosmology}

\quad The scalar theory defined by the Lagrangian \eqref{lagrangian} also finds application in models of cosmological inflation. Thus, a typical action in the Jordan frame can be rewritten in the following form:
\beq
S_J=\int d^4x \sqrt{-g} \left( - \dfrac{1}{2} F(\phi) R + \dfrac{1}{2} \delta_{ab} g_{\mu \nu}\partial^\mu\phi^a\partial^\nu\phi^b - \textbf{V}(\phi)\right),
\eeq
where $g=\det (g_{\mu \nu})$ and the function $F(\phi)$ is defined by:
\beq
F(\phi) = 1  + \dfrac{2}{R} \textbf{W}(\phi).
\eeq
Here, the quantum corrections are taken into account via the effective potential parts $\textbf{V}(\phi)$ and $\textbf{W}(\phi)$ in \eqref{series}.

The equations of motion and Friedmann equation in the slow-roll limit in the Jordan frame are given as:
\beq
3 H \dot{\phi}^a \simeq - F^2 \delta^{ab} \tilde{\textbf{V}}_{,b}\ ,
\label{eomJ}
\eeq
\beq
3 H^2 \simeq F \tilde{\textbf{V}}\ ,
\label{FeJ}
\eeq
where $\tilde{\textbf{V}}(\phi)={\textbf{V} (\phi)}/{F^2}$.

Under a conformal transformation of the metric tensor
\beq
\tilde{g}_{\mu \nu} =  F\left( \phi\right) g_{\mu \nu},
\eeq
we can transform into an action in the Einstein frame:
\beq
S_E =\int d^4x \sqrt{-\tilde{g}} \left(- \dfrac{\tilde{R}}{2} + \dfrac{1}{2} \mathcal{K}_{ab} \tilde{g}^{\mu \nu} \partial_\mu \phi^a \partial_\nu \phi^b - \tilde{\textbf{V}}(\phi)\right),
\label{S_E}
\eeq
where
\beq
\mathcal{K}_{ab}(\phi) = \frac{1}{F} \left(\delta_{ab} + \dfrac{3 F_{,a} F_{,b} } {2 F} \right),
\eeq
and the total derivative term is thrown away. 

In case with $N=1$ the considered action \eqref{S_E} can be reduced to an action with a canonically normalised kinetic term by introducing the scalar field $\tilde{\phi}$. For the classical case $F = 1 + \xi \phi^2 $ in the limit $\xi \phi^2 \gg 1$, which is realised in inflationary cosmology, we obtain:
\beq
\dfrac{d \tilde{\phi}}{d \phi} = \dfrac{ \sqrt{6} \xi \phi}{1 + \xi \phi^2}
\eeq
the solution of which is
\beq
\tilde{\phi} = \sqrt{\dfrac{3}{2}}\ln\left( 1 + \xi \phi^2 \right).
\eeq
Thus, the classical case of the power-like potential \eqref{V_0} in the Einstein frame takes the following form:
\beq
\tilde{\textbf{V}}(\tilde{\phi}) = \dfrac{\la}{4! \xi^{\frac{p}{2}}} e^{ -2 \sqrt{\frac{2}{3}} \phi} \left( e^{ \sqrt{\frac{2}{3}} \phi} - 1 \right)^{\frac{p}{2}}.
\label{V_tilde}
\eeq
It is known that $R + R^{2 \hat{p}}$ inflation leads to a similar type of potential \cite{Martin:2013tda, Felice:2010}. The relationship between the indices of the models is  $\hat{p} = \frac{p}{2(p-2)}$. For example, for $\hat{p} = 1$, one recovers the Starobinsky inflationary model \cite{Starobinsky:1980}. Indeed, the considered case with $\hat{p} = 1$ corresponds to $p=4$ in \eqref{V_tilde}:
\beq
\tilde{\textbf{V}}(\tilde{\phi}) \simeq \left(1 - e^{ - 2 \sqrt{\frac{2}{3}} \phi} \right)^{2}.
\eeq
Alternative parameterizations may also be considered \cite{Odintsov:2025eiv}.

Within the slow-roll limit, the equations of motion and the Friedmann equation in the Einstein frame take the form:
\beq
3 \tilde{H} \dfrac{d \phi ^ a}{d \tilde{t}} = \mathcal{K}^{ab} \tilde{\textbf{V}}_{,b}\ ,
\label{eomE}
\eeq
\beq
3 \tilde{H}^2 = \tilde{V},
\label{FeE}
\eeq
where $\tilde{H}$ is the Hubble parameter in the Einstein frame, and $d\tilde{t} = \sqrt{F} dt$ is proper time.

For instance, starting from the definition of the number of $e$-folds $\mathcal{N}$ and \eqref{eomJ}-\eqref{FeJ}, one can obtain 
\beq
\mathcal{N}  \simeq  \intop_{\phi_{\mathrm{end}}}^{\phi} \dfrac{\tilde{\textbf{V}}}{F} \dfrac{\tilde{\textbf{V}}_{,a}}{\delta^{lk} \tilde{\textbf{V}}_{,l} \tilde{\textbf{V}}_{,k} } d \phi^a.
\label{N_e_a}
\eeq
In turn, the scalar spectral index $n_s$ can be written as \cite{White:2013ufa}:
\beq
n_s - 1 \simeq 2 \dfrac{ {d \tilde{H}}/{d \tilde{t}} }{ \tilde{H}^2 } - \dfrac{2}{\mathcal{N}_{,a}\mathcal{N}^{,a}} + \dfrac{ 2 \mathcal{N}^{,a}\mathcal{N}^{,b} }{ 3 \tilde{H}^2 \mathcal{N}_{,k}\mathcal{N}^{,k} } \tilde{\nabla}_a \tilde{\nabla}_b \tilde{\textbf{V}}\ ,
\label{ns_ab}
\eeq
where we neglect the curvature term $\tilde{R}_{adbc} \frac{d \phi^c}{d\tilde{t}}\frac{d \phi^d}{d\tilde{t}} \ll \mathcal{O}(\epsilon)$.
Here, $\tilde{\nabla}_a$ is the covariant derivative with respect to the Einstein frame ﬁeld-space metric $\mathcal{K}_{ab}$, where the
Christoffel symbols are:
\beq
\mathcal{K}_{ad} \,^{(\mathcal{K})}\Gamma^d_{bc} = - \frac{1}{2F^2}\left(\delta_{ab} F_{,c} + \delta_{ac} F_{,b} - \delta_{bc} F_{,a} \right) - \frac{3 F_{,a} F_{,b} F_{,c}}{2 F^3}+\frac{3 F_{,a} F_{,bc}}{2F^2}.
\label{Csym}
\eeq

Using \eqref{eomE}, \eqref{FeE}, \eqref{N_e_a} and \eqref{Csym}, one can rewrite an expression \eqref{ns_ab} for $n_s$:
\beq
n_s - 1 \simeq - 6 \epsilon_{\scriptscriptstyle\tilde{V}} + 2 \eta_{\scriptscriptstyle\tilde{V}},
\label{ns_final}
\eeq
where we define the slow-roll parameters as:
\beq
\epsilon_{\scriptscriptstyle\tilde{V}} = \dfrac{F}{2} \dfrac{ \delta^{ab} \tilde{\textbf{V}}_{,a} \tilde{\textbf{V}}_{,b}}{ \tilde{\textbf{V}}^2}\ ,
\label{eps_ab}
\eeq
\beq
\eta_{\scriptscriptstyle\tilde{V}} = F \dfrac{\tilde{\textbf{V}}^{,a} \tilde{\textbf{V}}^{,b}}{\tilde{\textbf{V}}_{,e} \tilde{\textbf{V}}^{,e}} \dfrac{\tilde{\textbf{V}}_{,ab}}{\tilde{\textbf{V}}} + \dfrac{F_{,a}}{2} \dfrac{ \tilde{\textbf{V}}^{,a} }{ \tilde{\textbf{V}}}\ .
\label{eta_ab}
\eeq

It should be noted that all quantities depend on the modulus $\phi = \sqrt{\phi_a \phi_a}$. It is also useful to give field derivatives of a quantity $A(\phi)$ with respect to $\phi_a, \phi_b$:
\beq
A_{,a}(\phi) = A_{,\phi}(\phi) \dfrac{\phi_a}{\phi},
\label{A_a}
\eeq
\beq
A_{,a b} = A_{,\phi\phi}(\phi) \dfrac{\phi_a \phi_b}{\phi^2} + 2 \dfrac{\partial A(\phi)}{\partial (\phi^2)} \left( \delta_{ab} - \dfrac{\phi_a \phi_b}{\phi^2} \right)
\label{A_ab}
\eeq
One can notice that this has already been applied in formula \eqref{D_inv_ab}.
Taking this remark into account, the convolutions in \eqref{eps_ab} and \eqref{eta_ab} are simplified:
\beq
\epsilon_{\scriptscriptstyle\tilde{V}} = \dfrac{F}{2} \left( \dfrac{\tilde{\textbf{V}}_{,\phi}}{\tilde{\textbf{V}}} \right)^2,
\label{eps_final}
\eeq
\beq
\eta_{\scriptscriptstyle\tilde{V}} = F \dfrac{ \tilde{\textbf{V}}_{,\phi\phi}}{\tilde{\textbf{V}}} + \dfrac{F_{,\phi}}{2} \dfrac{  \tilde{\textbf{V}}_{,\phi} }{ \tilde{\textbf{V}}},
\label{eta_final}
\eeq
which coincides with the single-field result for a non-minimal coupling to gravity derived in \cite{Karciauskas:2022jzd}. When describing the background evolution in an $SO(N)$ symmetric inflationary model, the multi-field problem reduces to a single-field slow-roll along the radial component $|\phi|$. Naturally, to study fluctuations around the background, it is necessary to study the mechanisms of suppression of the isocurvature modes by generating heavy masses using more sophisticated potentials \cite{White:2013ufa}. However, we address this issue to the future publications. Similarly, we can also present the expression for the tensor-to-scalar ratio $r$:
\beq
r  = 16 \epsilon_{\scriptscriptstyle\tilde{V}}.
\label{r_final}
\eeq

\begin{figure}[ht]
    \begin{center}
    \begin{minipage}{0.49\textwidth}
            \epsfxsize=8.3cm
            \epsffile{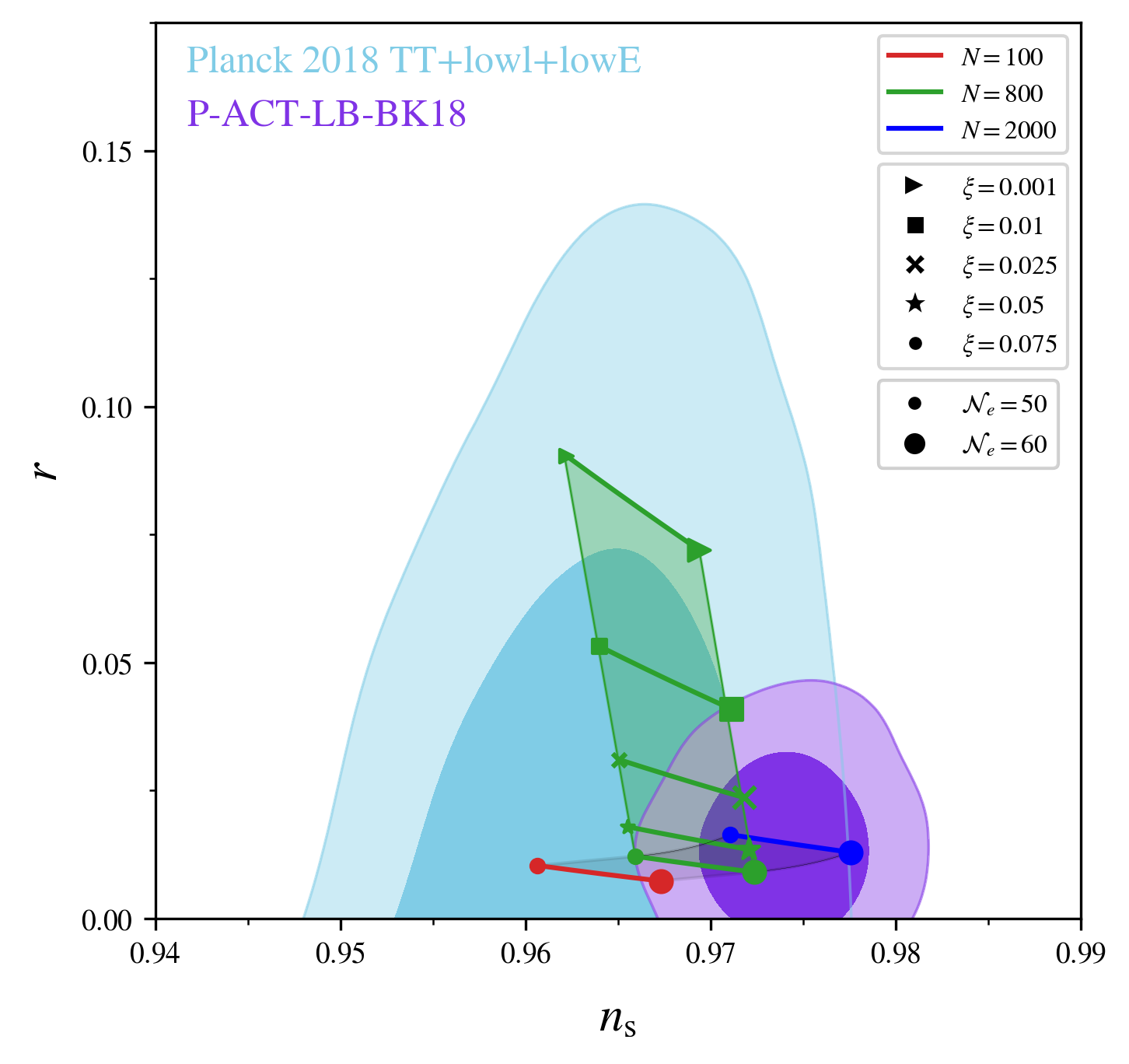}
            \label{fig::ns_r_R_p4}
        \end{minipage}
        \hfill
        \begin{minipage}{0.49\textwidth}
            \epsfxsize=8.3cm
            \epsffile{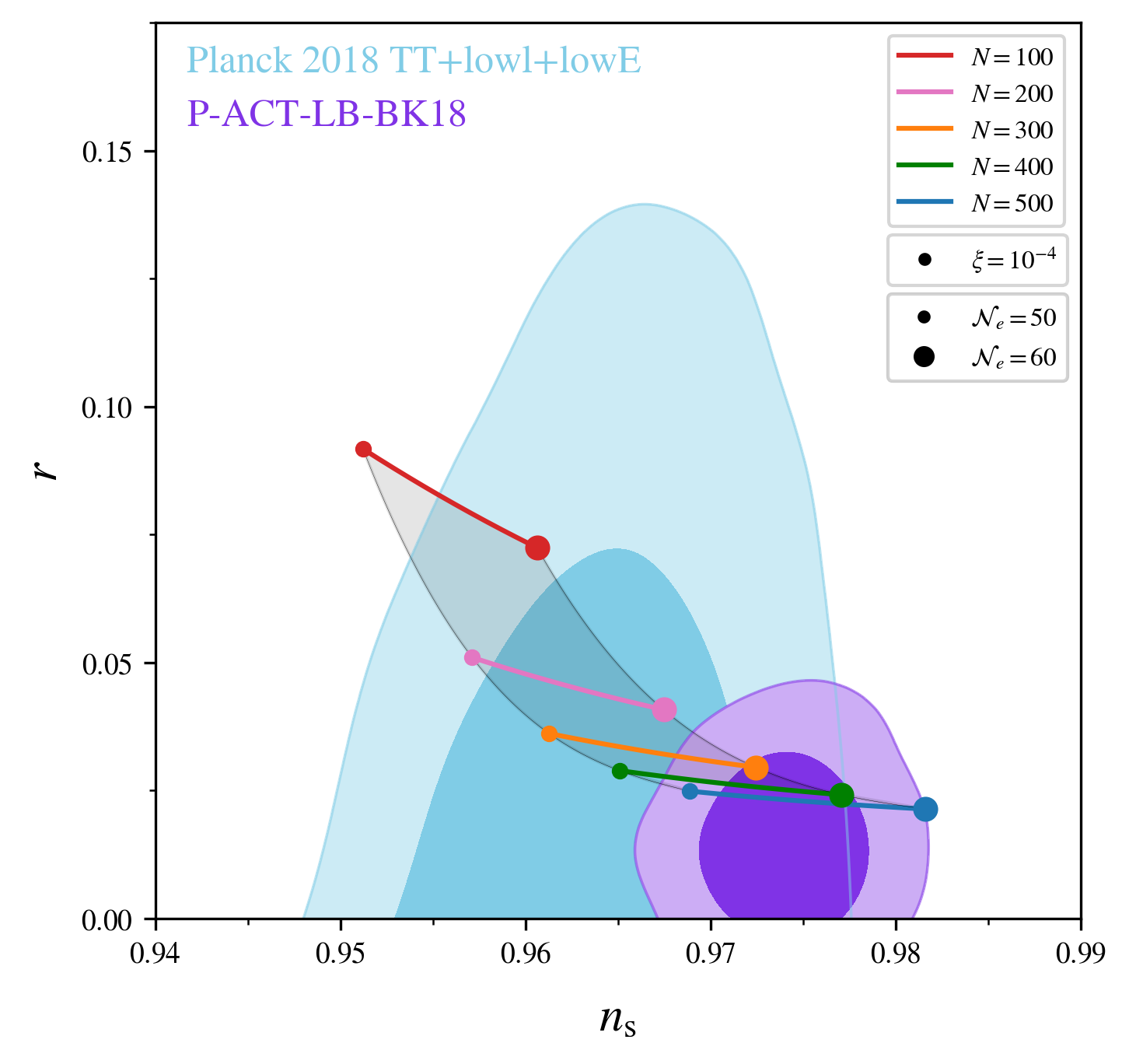}
            \label{fig::ns_r_R_p6}
        \end{minipage}
    \end{center}
    \vspace{-0.9cm}
    \caption{The $(n_s, r)$-diagram for effective potentials with $p=4$ (left), $p=6$ (right) and observed data.}
    \label{fig::ns_r_R_p4_p6}
\end{figure}

Using expressions \eqref{ns_final} and \eqref{r_final}, we evaluate the cosmological parameters for the considered models of power-like potentials with $p=4$ and $p=6$. The parametric plot of $n_s$ versus $r$, compared with the Planck 2018 data \cite{Planck:2018jri} and the combined P-ACT-LB-BK dataset \cite{ACT:2025blo, BICEP:2021xfz}, is shown in Fig.~\ref{fig::ns_r_R_p4_p6}. Larger values of the parameter $\xi$ lead to smaller $r$ and brings the predictions into agreement with the observational constraints. For the case $p=4$, this is illustrated by the green region with various markers. Furthermore, varying $N$ produces a shift along the $n_s$ direction. This behavior is illustrated by the gray region, within which the individual values of $N$ are distinguished by different colors. Such flexible behavior allows both models to be consistent with either the Planck or the P-ACT-LB-BK observational constraints by an appropriate choice of $N$.

\section{Conclusion}
\quad We have derived the RG equation in a general approach that enables one to calculate quantum corrections to an arbitrary classical potential in $SO(N)$ symmetric scalar theories within the leading logarithmic approximation, taking into account the non-minimal coupling to gravity. We analysed cases of power-like potentials with $p=4$ and $p=6$. In accordance with the results obtained in \cite{Ishikawa:1983kz}, for the power-law potential with $p=4$, regimes with critical curvature values $R_{C_i}$ were identified. For the case with $p=6$, the behaviour of the effective potential was examined for different values of the model parameters. We also showed that the obtained effective potentials of $SO(N)$ symmetric models can be applied in the theory of cosmological inflation. The cosmological parameters $n_s$ and $r$ were computed for both models and compared with the Planck 2018 and P-ACT-LB-BK observational data. It was demonstrated that varying the parameter $N$ allows agreement with the confidence region of both datasets. 

Interestingly, the regime with a locally flat plateau of the potential arising from the emergence of an additional minimum may be of interest for explaining the mechanism of primary black hole formation \cite{Ivanov:1994, Ezquiaga:2018, Ballesteros:2018, Frolovsky:2023}. Exploring this regime under inflationary constraints and comparing it with observational bounds on PBH production is one of the important directions for the work in progress.

\subsection*{Acknowledgements}
\quad Iakhibbaev Ravil and Tolkachev Denis thank the CERN Theoretical Department for their hospitality during the preparation of this work. The authors are also grateful to D.I. Kazakov for fruitful discussions and valuable comments.

\bibliographystyle{unsrt}
\bibliography{refs}

\begin{thebibliography}{10}

\bibitem{Coleman:1973jx}
Sidney~R. Coleman and Erick~J. Weinberg.
\newblock {Radiative Corrections as the Origin of Spontaneous Symmetry Breaking}.
\newblock {\em Phys. Rev. D}, 7:1888--1910, 1973.

\bibitem{Jackiw:1974cv}
R.~Jackiw.
\newblock {Functional evaluation of the effective potential}.
\newblock {\em Phys. Rev. D}, 9:1686, 1974.

\bibitem{Kastening:1991gv}
Boris~M. Kastening.
\newblock {Renormalization group improvement of the effective potential in massive $\phi^4$ theory}.
\newblock {\em Phys. Lett. B}, 283:287--292, 1992.

\bibitem{Ford:1991hw}
C.~Ford and D.~R.~T. Jones.
\newblock {The Effective potential and the differential equations method for Feynman integrals}.
\newblock {\em Phys. Lett. B}, 274:409--414, 1992.
\newblock [Erratum: Phys.Lett.B 285, 399 (1992)].

\bibitem{ChungThreeLoop:1998}
J.~M. Chung and B.~K. Chung.
\newblock Three-loop effective potential of o(n) $\phi^4$ theory, 1998.

\bibitem{Buchbinder:1987jf}
I.~L. Buchbinder and S.~D. Odintsov.
\newblock {Asymptotical behavior of the effective potential of the composite fields in curved space-time}.
\newblock {\em EPL}, 4:147--152, 1987.

\bibitem{Buchbinder:2017lnd}
I.~L. Buchbinder, S.~D. Odintsov, and I.~L. Shapiro.
\newblock {\em {Effective Action in Quantum Gravity}}.
\newblock Routledge, 10 2017.

\bibitem{Filippov:2025}
V.~A. Filippov, R.~M. Iakhibbaev, and D.~M. Tolkachev.
\newblock All-loop effective potential for arbitrary scalar models in curved space-time.
\newblock {\em Classical and Quantum Gravity}, 42(14):145006, July 2025.

\bibitem{Kazakov:2022pkc}
D.~I. Kazakov, R.~M. Iakhibbaev, and D.~M. Tolkachev.
\newblock {Leading all-loop quantum contribution to the effective potential in general scalar field theory}.
\newblock {\em JHEP}, 04:128, 2023.

\bibitem{Kazakov:2023tii}
D.~I. Kazakov, R.~M. Iakhibbaev, and D.~M. Tolkachev.
\newblock {Leading all-loop quantum contribution to the effective potential in the inflationary cosmology}.
\newblock {\em JCAP}, 09:049, 2023.

\bibitem{BP}
N.~N. Bogoliubow and O.~S. Parasiuk.
\newblock {\"U}ber die multiplikation der kausalfunktionen in der quantentheorie der felder.
\newblock {\em Acta Mathematica}, 97:227--266, 1957.

\bibitem{Hepp}
K.~Hepp.
\newblock {Proof of the Bogolyubov-Parasiuk theorem on renormalization}.
\newblock {\em Commun. Math. Phys.}, 2:301--326, 1966.

\bibitem{Zimmermann}
W.~Zimmermann.
\newblock {Convergence of Bogolyubov's method of renormalization in momentum space}.
\newblock {\em Commun. Math. Phys.}, 15:208--234, 1969.

\bibitem{Planck:2018jri}
Y.~Akrami et~al.
\newblock {Planck 2018 results. X. Constraints on inflation}.
\newblock {\em Astron. Astrophys.}, 641:A10, 2020.

\bibitem{ACT:2025blo}
Thibaut Louis et~al.
\newblock {The Atacama Cosmology Telescope: DR6 power spectra, likelihoods and {\ensuremath{\Lambda}}CDM parameters}.
\newblock {\em JCAP}, 11:062, 2025.

\bibitem{BICEP:2021xfz}
P.~A.~R. Ade et~al.
\newblock {Improved Constraints on Primordial Gravitational Waves using Planck, WMAP, and BICEP/Keck Observations through the 2018 Observing Season}.
\newblock {\em Phys. Rev. Lett.}, 127(15):151301, 2021.

\bibitem{Ishikawa:1983kz}
Kiyoshi Ishikawa.
\newblock {Gravitational effect on effective potential}.
\newblock {\em Phys. Rev. D}, 28:2445, 1983.

\bibitem{Bunch:1979uk}
T.~S. Bunch and L.~Parker.
\newblock {Feynman Propagator in Curved Space-Time: A Momentum Space Representation}.
\newblock {\em Phys. Rev. D}, 20:2499--2510, 1979.

\bibitem{Birrell:1982ix}
N.~D. Birrell and P.~C.~W. Davies.
\newblock {\em {Quantum Fields in Curved Space}}.
\newblock Cambridge Monographs on Mathematical Physics. Cambridge University Press, Cambridge, UK, 1982.

\bibitem{Petrov:1969}
A.Z. Petrov.
\newblock {\em {Einstein Space}}.
\newblock Pergamon, Oxford, 1969.

\bibitem{Sobreira:2011ep}
Flavia Sobreira, Baltazar~J. Ribeiro, and Ilya~L. Shapiro.
\newblock {Effective Potential in Curved Space and Cut-Off Regularizations}.
\newblock {\em Phys. Lett. B}, 705:273--278, 2011.

\bibitem{Iakhibbaev:2024fjf}
R.~M. Iakhibbaev and D.~M. Tolkachev.
\newblock {Effective potential in leading logarithmic approximation in non-renormalisable $SO(N)$ scalar field theories}.
\newblock {\em Int. Mod. Phys. J. A}, 2025.

\bibitem{BogoliubovBook}
N.~N. Bogoliubov and D.V. Shirkov.
\newblock {\em {Introduction To The Theory Of Quantized Fields}}.
\newblock Nauka, Moscow, 1957.
\newblock English transl.: {\it{ Introduction to the Theory of Quantized Fields, 3rd ed.}}, New York, Wiley, 1980.

\bibitem{Ivanov:1994}
P.~Ivanov, P.~Naselsky, and I.~Novikov.
\newblock Inflation and primordial black holes as dark matter.
\newblock {\em Phys. Rev. D}, 50:7173--7178, Dec 1994.

\bibitem{Ezquiaga:2018}
Jose~María Ezquiaga, Juan García-Bellido, and Ester Ruiz~Morales.
\newblock Primordial black hole production in critical higgs inflation.
\newblock {\em Physics Letters B}, 776:345–349, January 2018.

\bibitem{Ballesteros:2018}
Guillermo Ballesteros and Marco Taoso.
\newblock Primordial black hole dark matter from single field inflation.
\newblock {\em Physical Review D}, 97(2), January 2018.

\bibitem{Frolovsky:2023}
Daniel Frolovsky and Sergei~V. Ketov.
\newblock Fitting power spectrum of scalar perturbations for primordial black hole production during inflation.
\newblock {\em Astronomy}, 2(1):47--57, 2023.

\bibitem{Martin:2013tda}
Jerome Martin, Christophe Ringeval, and Vincent Vennin.
\newblock {Encyclop\ae{}dia Inflationaris}: {Opiparous Edition}.
\newblock {\em Phys. Dark Univ.}, 5-6:75--235, 2014.

\bibitem{Felice:2010}
Antonio De~Felice and Shinji Tsujikawa.
\newblock f(r) theories.
\newblock {\em Living Reviews in Relativity}, 13(1), June 2010.

\bibitem{Starobinsky:1980}
A.A. Starobinsky.
\newblock A new type of isotropic cosmological models without singularity.
\newblock {\em Physics Letters B}, 91(1):99--102, 1980.

\bibitem{Odintsov:2025eiv}
S.~D. Odintsov and V.~K. Oikonomou.
\newblock {Power-law F(R) gravity as deformations to Starobinsky inflation in view of ACT}.
\newblock {\em Phys. Lett. B}, 870:139907, 2025.

\bibitem{White:2013ufa}
Jonathan White, Masato Minamitsuji, and Misao Sasaki.
\newblock {Non-linear curvature perturbation in multi-field inflation models with non-minimal coupling}.
\newblock {\em JCAP}, 09:015, 2013.

\bibitem{Karciauskas:2022jzd}
Mindaugas Kar\v{c}iauskas and Jos\'e Jaime~Terente D\'\i{}az.
\newblock {Slow-roll inflation in the Jordan frame}.
\newblock {\em Phys. Rev. D}, 106(8):083526, 2022.

\end{thebibliography}
\end{document}